# Angle-Independent Plasmonic Substrates for Multi-Mode Vibrational Strong Coupling with Molecular Thin Films

*Zachary T. Brawley [1], S. David Storm [2,†], Diego A. Contreras Mora[3], Matthew Pelton[2], Matthew Sheldon[1,3,*]*

[1]Department of Materials Science and Engineering, Texas A&M University, College Station, TX, USA

[2]Department of Physics, UMBC (University of Maryland, Baltimore County), Baltimore, MD, USA

[3]Department of Chemistry, Texas A&M University, College Station, TX, USA

[†]Current address: Department of Physics and Astronomy, Vanderbilt University, Nashville, TN, USA

*Corresponding Author: sheldonm@tamu.edu

*Abstract*

Vibrational strong coupling of molecules to optical cavities based on plasmonic resonances has been explored recently, because plasmonic near-fields can provide strong coupling in sub-diffraction limited volumes. Such field localization maximizes coupling strength, which is crucial for modifying the vibrational response of molecules and, thereby, manipulating chemical reactions. Here, we demonstrate an angle-independent plasmonic nanodisk substrate that overcomes limitations of traditional Fabry–Pérot optical cavities, because the design can strongly couple with all molecules on the surface of the substrate regardless of molecular orientation. We demonstrate that the plasmonic substrate provides strong coupling with the C=O vibrational stretch of deposited films of PMMA. We also show that the large linewidths of the plasmon resonance allows for simultaneous strong coupling to two, orthogonal water symmetric and asymmetric vibrational modes in a thin film of copper sulfate monohydrate deposited on the substrate surface. A three-coupled-oscillator model is developed to analyze the coupling strength of the plasmon resonance with these two water modes. With precise control over the nanodisk diameter, the plasmon resonance is tuned systematically through the modes, with the coupling strength to both modes varying as a function of the plasmon frequency, and with strong coupling to both modes achieved simultaneously for a range of diameters. This work may aid further studies into manipulation of the ground-state chemical landscape of molecules by perturbing multiple vibrational modes simultaneously and increasing the coupling strength in sub-diffraction limited volumes.

*Introduction*

Strong coupling of molecular vibrational modes to optical (infrared) cavities has been explored increasingly in recent years as an avenue to modify the intrinsic vibrational response of molecular systems and, thereby, the chemical properties associated with the vibrating bond.[1–5] When molecules are inside an optical cavity, the two systems can coherently exchange radiative energy if the frequency of the cavity resonance is tuned to the molecular vibrational mode. When the exchange of energy occurs faster than the damping rates of the cavity and of the vibration, the system is said to be in the strong coupling regime.[6] Energy splitting of the original eigenstates, otherwise known as Rabi splitting, gives rise to hybrid polariton modes with higher and lower frequencies than the original resonance, with consequent modification of the bond energy.[3]

Optical cavities based on Fabry–Pérot (F.P.) resonances have been the primary optical platform used to promote vibrational strong coupling to date, due to their high quality factor, $Q$. Recently, this platform has helped to increase understanding of the underlying dynamics of polariton chemistry, such as relaxation lifetimes,[7] energy transfer,[8] and the role of density of states on polariton-modified reactions.[9] However, even though F.P. geometries are useful due to their low loss, they have intrinsic limitations, and overcoming these is the focus of this report. Because the magnitude of vibrational strong coupling is directly proportional to both the number of molecules present *and* the dot product between the electric field vector of the cavity and the vibrational dipole moment,[3,4,10] coupling with an entire ensemble of molecules cannot be achieved unless all of the vibrating bonds are aligned with the electric field inside the cavity. Another limitation of F.P. cavities is that the optical field concentration is diffraction limited. Prior work has shown that coupling strength is inversely proportional to the square root of the electric field mode volume.[11,12] Physically, this means that the highest degree of coupling occurs when the electric field density is largest, or equivalently, when the mode is volume is as small as possible. In F.P. cavities, the maximum electric field density is constrained by the wavelength-scale limit on the minimum mode volume.

As an alternative, plasmonic metal nanomaterials are quickly gaining attention as a new strategy for vibrational strong coupling.[13,14] Although plasmonic devices are lossy in comparison with F.P. resonances, they may overcome their lower Q-factors and out-compete damping processes by taking advantage of the intense optical near-fields at the metal surface, which can enhance and concentrate light by many orders of magnitude in sub-diffraction-limited volumes.[15]

This near-field concentration is created by an external optical field driving the free electrons in the metal into a collective oscillation, which leads to a localized surface plasmon resonance (LSPR). Further, plasmons can be tailored to exhibit resonances through the visible and mid infrared (IR) spectrum, and with much greater flexibility regarding the suitable molecular dipoles that can couple with them (i.e. lower momentum matching constraints), compared with cavities based on far-field optical resonances. The enhanced fields created in sub-diffraction-limited volumes suggest the plasmonic devices can allow for higher coupling strengths than F.P. cavities, overcoming their broader resonant linewidths (i.e. higher damping rates), in order to enable multi-mode strong coupling as we describe below. While there have been several studies coupling excitons produced by quantum dots,[16–18] J-aggregates,[19–24] semiconductors,[25] or molecular transitions[26–28] to plasmons, there is significantly less work that has studied vibrational coupling of bulk molecular systems to plasmonic structures.[29,30]

In this study, we developed a plasmonic nanodisk substrate with a resonance that is tunable throughout the near-IR spectrum and that exhibits no angle dispersion. That is, the plasmonic mode couples equally with all dipole orientations. We used a Salisbury screen design,[31] which takes advantage of constructive interference in the dielectric layer of a metal-insulator-metal to create a an absorber mode that, in combination with the LSPR from the plasmonic nanodisks, also show angle-independent, near-unity IR absorptivity and emissivity[32] (Fig. 1). Prior work has shown that these geometries have angle-independent resonances that can be tuned in the IR regime.[33–35] We hypothesized that these substrates may be an ideal platform for studying vibrational strong coupling, because (1) deposited films on the substrate surface can couple to the plasmon regardless of molecular orientation, (2) coupling occurs in sub-diffraction limited volumes, maximizing the coupling strength in the system.

Additionally, because of the broad linewidth of the plasmonic mode, the substrates allow for strong coupling to multiple vibrational modes simultaneously. This is in contrast to F.P. cavities which usually have narrow optical modes (i.e. low damping rates). While a narrow F.P. mode is useful for strong coupling due to the low loss of the system, it limits the cavity's ability to couple to multiple, closely spaced vibrational modes. Instead, previous studies with F.P. cavities have shown coupling to spectrally isolated modes taking advantage of multiple optical resonances,[36,37] while others have shown polariton hybridization of different molecular species with similar spectral frequencies.[38] Related work from Menghrajani *et al.* demonstrated coupling of three different vibrational modes in PMMA to both a F.P. cavity and a plasmonic grating, although only one polariton mode reached the strong coupling regime.[39] Building from these previous reports, this study shows how a plasmonic substrate can strongly couple to two, orthogonal vibrational modes simultaneously, potentially opening the door for more sophisticated coherent energy transfer between a cavity and chemical system.

To study multi-mode strong coupling effects, copper sulfate monohydrate [$CuSO_4(H_2O)_1$] was deposited as a thin film on the substrate surface. This molecule was analyzed because it is easy to characterize using Raman and IR spectroscopy and has strong, spectrally isolated vibrational modes in the mid IR regime.[40,41] Copper sulfate is also particularly interesting because the most prominent absorption band near 3200 $cm^{-1}$ is due to two distinct, orthogonal symmetric and asymmetric stretching modes of the associated water molecule, giving insight into how the plasmonic substrate may couple to both modes at room temperature. Recently, Vergauwe *et al.* showed that by strongly coupling to one water stretching mode with a F.P. cavity, the rate of an enzymatic hydrolysis reaction was modified.[42] We hypothesize that by strongly coupling to both water symmetric and asymmetric stretching modes simultaneously, the potential energy landscape of the molecule may be perturbed even more than what can be achieved with single mode coupling to a F.P. resonance, leading to increased modification of reaction rates. We note that in pure liquid water the symmetric and asymmetric stretching modes are not as well isolated as in $CuSO_4(H_2O)_1$.

We also analyzed the total hemispherical absorptivity of the substrate to demonstrate that the plasmon resonance is indeed angle-independent, suggesting that all molecules in the optical near-field can participate in strong coupling with the substrate. Finally, a three-coupled-oscillator model was developed that fits to the coupling data based on the assumption the plasmon couples to the two orthogonal water stretching modes simultaneously, due their close spectral spacing. We observed that, in general, the plasmon mode couples more strongly to the asymmetric water stretching mode, likely due to the stronger transition dipole of the that molecular vibration. Our study confirms that strong coupling can be obtained for the water symmetric and asymmetric stretching modes simultaneously due to the unique design of the plasmonic substrate.

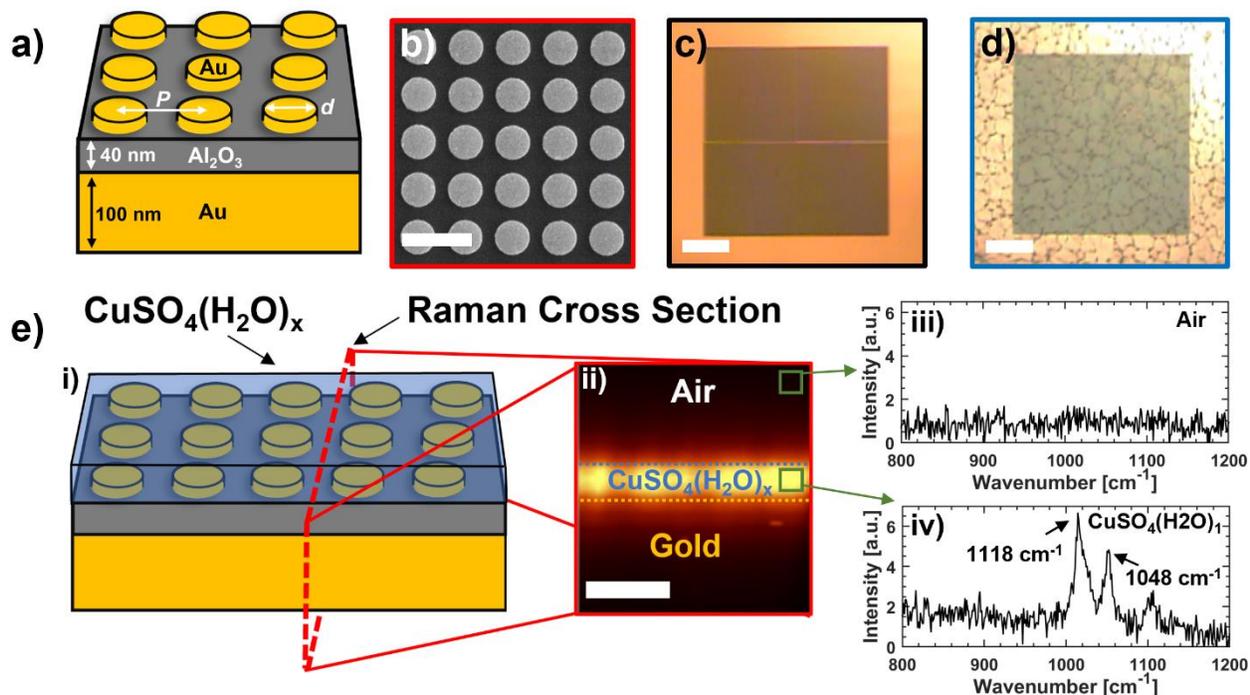

**FIG. 1.** Experimental design and fabrication of plasmonic coupling platform. (a) Schematic of the plasmonic substrate. The nanodisk height is held constant at 100 nm. The gap spacing is held constant

at 250 nm so that the pitch $P$ changes only as a function of nanodisk diameter $d$. (b) Scanning-electron-microscope image of a nanodisk array with $d = 680$ nm. The scale bar is 1 µm. (c) Optical image of the entire 200 × 200 µm nanodisk square array shown in (b). The scale bar is 20 µm. (d) The nanodisk array shown in (c) with 2 µm of $CuSO_4(H_2O)_1$ deposited on top. (e) Confocal Raman depth map of the plasmonic substrate with 2 µm of $CuSO_4(H_2O)_1$ deposited. (i) Schematic of plasmonic substrate with $CuSO_4(H_2O)_1$ on top. The dotted red lines show the Raman map region. (ii) The Raman depth map with horizontal dashed lines indicating the boundaries of the different regions. The scale bar is 4 µm. (iii) A spectrum obtained by the Raman map above the substrate. (iv) A spectrum obtained from the film region showing the prominent peaks of $CuSO_4(H_2O)_1$ at 1118 cm$^{-1}$ and 1048 cm$^{-1}$.

## *Methods*

### *Finite element method optimization of nanodisk substrate*

To determine the optical properties of the nanodisk substrate, full-wave optical simulations (finite element method (FEM), COMSOL) were used. Gold nanodisks with a tunable diameter and with a height of 100 nm were simulated on top of a 40 nm thick $Al_2O_3$ layer on top of an optically opaque smooth 100 nm gold substrate. A 5 nm chrome layer was placed between the nanodisks and the $Al_2O_3$ to model an adhesion layer used in fabrication. Air (n = 1) was used as the surrounding dielectric. Fig. 1a is a schematic of the substrate. Periodic boundary conditions were applied to simulate an infinite square array of gold nanodisks with a constant gap spacing of 250 nm in the x- and y-directions between adjacent nanodisks, meaning the pitch $P$ varied only as a function of the disk diameter $d$. The gold refractive index used was from Babar and Weaver[43] because it extends the refractive index from the visible to IR regime, and the refractive index for Al$_2$O$_3$ used was from Kischkat *et al.*[44] for similar reasons. The refractive index for chrome was obtained from Rakić *et al.*[45]

*Determining total hemispherical absorptivity*

The total hemispherical absorptivity of the structures was obtained from the FEM simulations by integrating the absorptance of the substrate from $\theta = 0°$ to $\theta = 80°$ and $\emptyset = 0°$ to $\emptyset = 90°$, where $\theta$ is the elevation angle from normal incidence and $\emptyset$ is the azimuthal angle. The symmetry of the square array implies that the other three hemispherical quadrants from $\emptyset = 90°$ to $\emptyset = 360°$ have the same angular absorptance as the first quadrant. The light was simulated as "unpolarized" by taking the average of S and P polarizations at every incident angle. The total hemispherical absorptivity was determined using the following equation:[46,47]

$$a(\lambda, \theta, \emptyset) = \frac{\int_0^\emptyset \int_0^\theta \int_{\lambda_1}^{\lambda_2} [1 - R(\lambda, \theta, \emptyset)] I_{BB}(\lambda) \cos\theta \, d\lambda \, d\Omega}{\int_0^\emptyset \int_0^\theta \int_{\lambda_1}^{\lambda_2} I_{BB}(\lambda) \cos\theta \, d\lambda \, d\Omega}$$

where $R(\lambda, \theta, \emptyset)$ is the reflectance of the substrate, $1 - R(\lambda, \theta, \emptyset)$ is the absorptance $a(\lambda, \theta, \emptyset)$ of the substrate, $I_{BB}$ is the spectral intensity of the energy absorbed by an ideal blackbody as a function of the wavelength, λ, in accordance with Planck's law, and $d\Omega = sin\theta d\theta d\emptyset$ is the solid angle. In our system, the transmittance of the substrate was zero due to the gold back reflector.

*Nanodisk substrate fabrication*

Once optimized nanodisk geometries were determined from the simulations, the structures were fabricated using electron-beam lithography. Base piranha and UV-ozone were used to clean a 1 cm × 1 cm silicon chip. A 100 nm gold layer was thermally evaporated (Lesker PVD electron-beam evaporator) onto the silicon chip. After the gold was deposited, a 40 nm layer of $Al_2O_3$ was deposited using RF sputtering (Lesker PVD RF sputterer). Electron beam lithography was used to fabricate 200 × 200 µm square nanodisk arrays, with each array's nanodisk diameters corresponding to the desired plasmon resonance. 950 PMMA A4 was used as the e-beam resist. Finally, a 5 nm chrome adhesion layer was deposited on top of the unfilled nanodisks developed into the resist, then a 100 nm gold layer was deposited on top of the chrome, which is the height of the nanodisks. Liftoff was performed in acetone using a combination of acetone pumping with a glass pipet and sonication. The resulting nanodisk array is shown in Fig. 1b (SEM) and Fig. 1c (optical). The array appears dark due to the absorbing nature of the plasmonic nanodisks in the visible regime.

*Thin film deposition*

$CuSO_4(H_2O)_5$ was mixed with nanopure water to form a 25 mM solution. The fabricated substrate was placed on a hot plate heated to ~310 ˚C. Once the substrate was up to temperature, the 25 mM solution was drop cast on top of the entire substrate, and the water was allowed to evaporate quickly, leaving behind only a $CuSO_4(H_2O)_1$ thin film. Due to the crystallinity of the $CuSO_4(H_2O)_1$, the thin film is a homogeneous, uniform layer. Thus, the thin film has an approximately constant number of molecules participating in the coupling across the surface region studies spectroscopically. An optical image of the deposited molecular thin film is shown in Fig. 1d. Notice the color change of the nanodisk array from Fig. 1c to 1d due to the change in refractive index of the surrounding medium.

To study coupling with a single vibrational mode, we spin coated 950 PMMA A4 at 1500 RPM to create a ~300 nm thin film on top of a new nanodisk substrate. A diameter, $d = 1200$ nm, was used in order to target the C=O stretch with the plasmon resonance, which is located at 1728 cm$^{-1}$ and is spectrally isolated from other vibrational modes.

*$CuSO_4(H_2O)_X$ thin film thickness*

A Witec RA300 confocal Raman microscope was used to determine the $CuSO_4(H_2O)_x$ thin film thickness. A 532 nm laser was directed through a 50× (numerical aperture 0.55) objective lens with an optical power of 360 µW. A Raman spatial scan, like that shown in Fig. e(i), was obtained by performing a depth profile map. The map started at z = +10 µm and moved in the -z direction in -1 µm increments to a final depth of z = -10 µm. The horizontal y-direction started at y = -10 µm and moved to y = +10 µm in +1 µm increments on each subsequent z-step of the depth profile. An integration time of 0.5 s was used on each step. The resulting cross-sectional depth map is shown in Fig. e(ii). The signal was filtered to select for the 1119 cm$^{-1}$ peak – the prominent spectral signature of $CuSO_4(H_2O)_1$ – indicating that the bright areas are showing only the signal from the $CuSO_4(H_2O)_1$.[48] By selecting for specific pixels in the depth profile, we observed that the dark area above the substrate is air [Fig. e(iii)] and the bright area on the substrate is a thin film of $CuSO_4(H_2O)_1$ [Fig. e(iv)], further supported by the prominent peaks at 1118 cm$^{-1}$ and 1048 cm$^{-1}$.[48] These Raman peaks show no signature of coupling because the plasmon resonant frequencies of our structures were made to range from 1500 cm$^{-1}$ – 4200 cm$^{-1}$, which is not resonant with the copper sulfate Raman peaks. Raman spectra of the molecules on and off the substrate are shown in Fig. S1 (Supplementary Information). From the depth map, the $CuSO_4(H_2O)_1$ thin film layer was determined to be ~2 µm thick.

*FTIR absorbance measurements*

A Shimadzu AIM-8800 automatic infrared microscope with a 20× objective was used to acquire all IR spectra at normal incidence. A 40 nm $Al_2O_3$ thin film on a 100 nm gold reflector was used as the background before acquiring spectra. The aperture was adjusted to acquire only the spectra from the nanodisks. Absorbance spectra were collected, which is the $Log_{10}$(Absorptance), and then the spectra were normalized to the highest spectral peak on a scale from 0 to 1. All data sets were averaged over 40 collected spectra at one position.

## *Results and Discussion*

For our substrate design, we have adapted the work of Abbas *et al*.,[33] which showed that silver nanodisks on $SiO_2$ with a silver back reflector can absorb nearly 100% of incident radiation, are IR tunable, and are angle-independent. Our computational analysis simulated the spectral absorptance versus the diameter of plasmonic gold nanodisks on top of $Al_2O_3$ with a gold back reflector. The gold was used rather than silver to reduce the effects of substrate oxidation, and the $Al_2O_3$ was used to avoid any mid-IR absorption that would occur with $SiO_2$. The results of the simulation (normal incidence) are shown in Fig. 2a. It is clear that the plasmonic resonance can be tuned across the entire mid-IR regime as a function of nanodisk diameter, and that the substrate absorbs nearly 100% of incident radiation at resonance. At larger diameters, a lower-order plasmon resonance is observed at higher wavenumbers, which is shown experimentally in Fig. S2a.

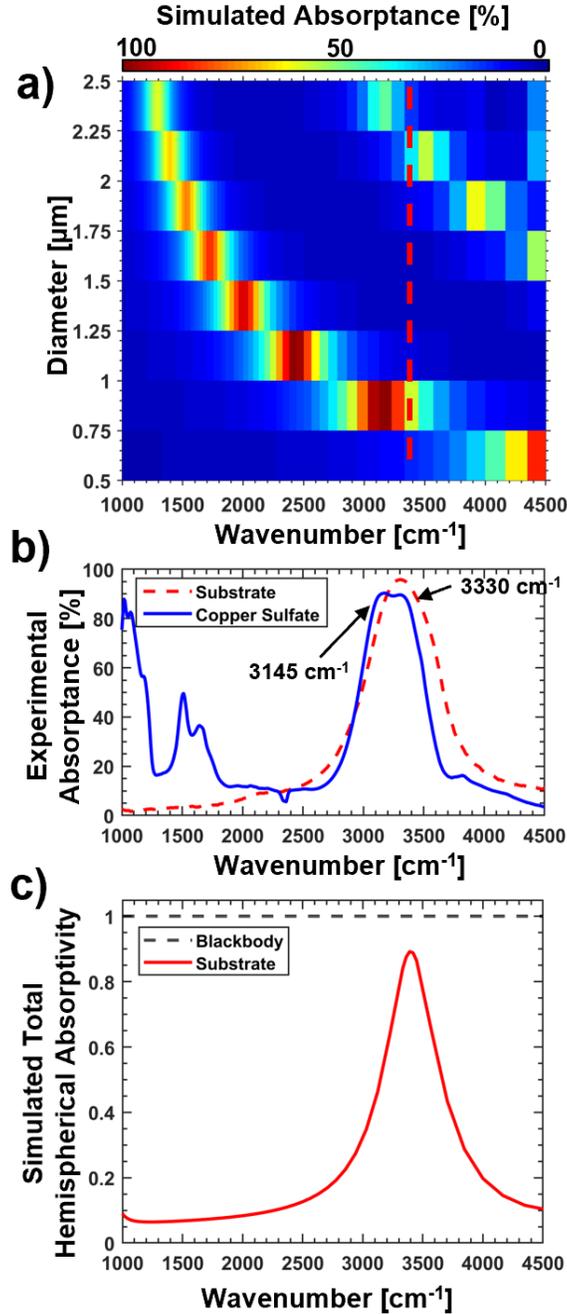

**FIG. 2.** Spectral tunability and angle independence of the plasmonic resonance. (a) Simulation (FEM) of the nanodisk array spectral absorptance as a function of diameter. The vertical red dashed line indicates the resonant frequency of 3333 cm$^{-1}$ for $d$ = 680 nm. (b) Corresponding experimental absorptance spectrum of the fabricated plasmonic substrate with $d$ = 680 nm (red dashed), and the absorptance spectrum of $CuSO_4(H_2O)_1$ (blue). The substrate is in resonance with both the symmetric and antisymmetric water stretching mode located at 3145 cm$^{-1}$ and 3330 cm$^{-1}$ respectively. (c) The calculated total hemispherical absorptivity of the nanodisk substrate from (b) in red compared to an ideal blackbody in black. The absorptance was integrated from $\theta = 0$ to $\theta = 80$ and from $\emptyset = 0$ to $\emptyset = 360$ over the entire hemisphere above the substrate.

Fig. 2b shows the experimentally measured IR absorptance spectrum of the $CuSO_4(H_2O)_1$ thin film on top of a 40 nm layer of $Al_2O_3$ on top of a gold back reflector. The dominant vibrational modes are the water bending mode of $CuSO_4(H_2O)_1$ near 1510 cm$^{-1}$ and the water symmetric and asymmetric stretching modes at 3145 cm$^{-1}$ and 3330 cm$^{-1}$, respectively.[40,41] Each of these vibrational modes in the mid-IR regime can be targeted individually by using the appropriate nanodisk diameter. In order to target the water stretching modes for these studies, a substrate was fabricated with a nanodisk diameter $d = 680$ nm, $Al_2O_3$ dielectric thin film layer with thickness $t = 40$ nm, and an optically thick Au back reflector. This geometry is indicated by the red dashed line in Fig. 2a. Fig. 2b also shows the experimental absorptance of that substrate geometry *without* $CuSO_4(H_2O)_1$ deposited (red dashed). The plasmon peak position was located at 3318 cm$^{-1}$, which targeted primarily the asymmetric water stretch. It is clear from the substrate spectrum that the resonance strongly concentrates and, thereby, absorbs nearly 100% of incident radiation over a narrow IR range, as required for strong coupling.

A crucial optical property of this substrate confirmed by computational analysis was the angle independence of the plasmonic resonance. To demonstrate this property, we simulated the total hemispherical absorptivity of the substrate.[46] That is, by integrating the spectral absorptance over the entire hemisphere, the frequencies in which the most radiation is absorbed over all angles was obtained. We observed that the substrate absorbs primarily only over a narrow frequency range and at the same spectral position as the experimentally obtained (normal incidence) spectrum, as shown by the red curve in Fig. 2c. The angle-independent plasmon resonance behavior of MIM nanodisks was also reported in previous work.[33,34] By normalizing the absorptivity to a perfectly absorbing blackbody spectrum (black dashed), we demonstrated that the nanodisk plasmon resonance is nearly as absorbing and as angle-independent (at resonance) as a blackbody, and that the damping rate, i.e. the peak width, is also angle-independent. By establishing this spectral design feature, we hypothesized that our system would demonstrate similar Rabi splitting as described in prior work.[19,24,27,49] Furthermore, due to the angle-independent resonance, nearly 100% of the molecules within the optical near-field could strongly couple to the plasmonic substrate, regardless of molecular orientation.

We next acquired spectra of various plasmonic substrates before and after a ~2 µm thin film layer of $CuSO_4(H_2O)_1$ was deposited on top of the substrate. The substrates were fabricated based on the simulation in Fig. 2a to tune the plasmon resonances systematically through the water stretching modes as a function of the nanodisk diameter. Fig. 3 shows the spectral positions of the plasmonic resonances without $CuSO_4(H_2O)_1$ on them (red dashed), with each individual panel showing a separate substrate. Diameters ranged from $d = 560$ nm to $d = 760$ nm. Six substrates were fabricated that tuned the plasmon resonance from 4274 cm$^{-1}$ to 2994 cm$^{-1}$. After the bare substrate spectra had been obtained, the $CuSO_4(H_2O)_1$ thin film was deposited on the surface. The spectra of the $CuSO_4(H_2O)_1$ on top of only the $Al_2O_3$ and gold back reflector is illustrated by the blue solid lines in Fig. 3. The prominent peaks at 3145 cm$^{-1}$ and 3330 cm$^{-1}$ are the same symmetric and asymmetric water stretches observed in Fig. 2a, only this time the spectra were normalized by the maximum. Notice that the thin-film peak locations are located at the same spectral positions in all six panels, meaning that each deposited salt film is in the same hydration state.

One important design feature to consider about the plasmonic substrates is that the resonance peak position red shifts as a function of the surrounding medium's refractive index. When the surrounding medium was air (n = 1), the red dashed spectra were obtained, illustrated in Fig. 3. However, when the surrounding medium was $CuSO_4(H_2O)_1$, the resonance frequencies red shift. This is shown in Fig. 3 panels (i) and (vi). In all six panels, the red dashed curves show the original plasmon substrate without $CuSO_4(H_2O)_1$ on it, the blue curve shows the $CuSO_4(H_2O)_1$ when deposited on a control region of the substrate without plasmonic nanodisks, and the black curve shows the spectra of $CuSO_4(H_2O)_1$ on top of the nanodisk array, illustrated by Fig. 1d and 1e. From panels (i) and (vi), the black curve clearly shows the red shift in the plasmonic substrate due to change in refractive index. If we focus on panel (i), the lineshape of the $CuSO_4(H_2O)_1$ is completely retained, and the plasmon peak is clearly observed at 3880 cm$^{-1}$, corresponding to a red shift of -270 cm$^{-1}$. Similarly, in panel (vi), a red shift of -183 cm$^{-1}$ is observed from an original plasmon frequency of 2939 cm$^{-1}$ to 2756 cm$^{-1}$. From these peak shifts, in combination with full-wave simulations, the refractive index of $CuSO_4(H_2O)_1$ was extracted as a function of wavenumber, and the results are shown in Fig. S2 (Supplementary Information). The refractive indices range from *n* = 1.27 farther from the water modes to *n* = 2 near the modes, indicating that the plasmon red shifts more when the plasmon position is in closer resonance with the vibrational peaks. This trend corroborates the expected Lorentzian lineshape of the dielectric function close to a molecular absorption mode, as expected for a damped harmonic oscillator, and this trend is also confirmed in the more sophisticated three-coupled-oscillator fitting model described below.[45,50,51]

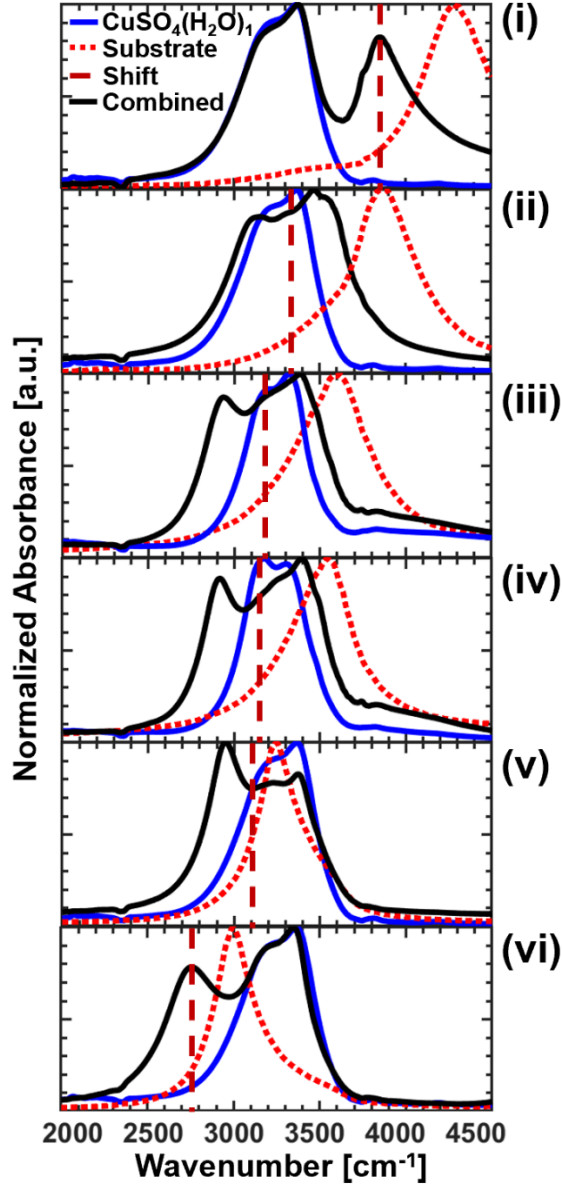

**FIG. 3.** The spectral absorbance (normalized) of a bare plasmonic substrate (red dashed), $CuSO_4(H_2O)_1$ thin film on $Al_2O_3$ with a gold back reflector (blue), and the combined spectra of the $CuSO_4(H_2O)_1$ thin film on top of the plasmonic substrate (black). The panels correspond to plasmon resonant frequencies (with approximate nanodisk diameters) at (i) 4274 cm$^{-1}$ ($d = $ ~560 nm), (ii) 3865 cm$^{-1}$ ($d = $ ~600 nm) (iii) 3595 cm$^{-1}$ ($d = $ ~670 nm) (iv) 3534 cm$^{-1}$ ($d = $ 680 nm), (v) 3248 cm$^{-1}$ ($d = $ ~720 nm), and (vi) 2994 cm$^{-1}$ ($d = $ ~760 nm). The vertical maroon dashed lines show the shifted plasmon resonance frequency when the surrounding medium is $CuSO_4(H_2O)_1$.

In Fig. 3 panels (ii) – (v), the plasmon resonances are not as easy to discern. Furthermore, the $CuSO_4(H_2O)_1$ symmetric and asymmetric water stretches are not clearly observed either. Instead, new line shapes are observed at a similar spectral position as the $CuSO_4(H_2O)_1$ modes, with a FWHM that is significantly broader than either the plasmon or $CuSO_4(H_2O)_1$ modes individually. In fact, panels (ii) – (v) show two characteristically different absorption features with peak positions at higher and lower frequencies than the original $CuSO_4(H_2O)_1$ modes, separated by a local minimum where the $CuSO_4(H_2O)_1$ frequencies were originally located. This behavior is clear indication of vibrational strong coupling.[6,52–54] When the plasmon and $CuSO_4(H_2O)_1$ spectrally overlap, coherent energy exchange is established between the two systems, causing two new hybrid polariton modes with higher and lower frequencies than the un-coupled modes. Thus, the plasmon resonance red shifts into resonance with the water vibrational modes, causing vibrational strong coupling to occur. The shifted plasmon resonance is depicted by the maroon vertical dashed lines in Fig. 3, as determined by the fitting equation discussed below.

The explanation for the changes in the observed broadening of the two polaritonic peaks as a function of the spectral position of the plasmon resonance can be understood by considering that there are two molecular modes present, the asymmetric and the symmetric water stretches. Several methods have been developed to account for coupling between molecular transitions and optical cavities including the transfer-matrix method, fitting matrix method, and full wave electrodynamic simulations.[52] Here, we adopted a semi-classical analytical approach from prior work, where the spectral lineshape for a single transition coupled to a single plasmon mode can be described quantitatively by modeling the transition (in this case, the molecular vibration) and the plasmon resonance as two coupled, classical harmonic oscillators (see Supplementary Information for details).[55–57] If we attempt to fit to this model, assuming only one molecular vibrational mode that is the mean value of the symmetric and asymmetric vibrational modes for $\omega_{vm}$, the fit deviates significantly from the experimental data. For example, Fig. 4a shows an attempt to fit this model to the data in panel (iv) in Fig. 3; the fit largely resembles the line shape of a single vibrational mode coupled to a plasmonic system established in prior work.[20,24,27,58] This suggests that the symmetric and asymmetric water stretches are not well described as one vibrational mode coupled with the substrate. Instead, it is important to consider both water stretching modes as separate modes that, while orthogonal in free molecules, can both interact with the plasmonic substrate. Similar multi-coupling spectra have previously been observed when coupling multiple vibrational modes of PMMA to a plasmonic grating.[39]

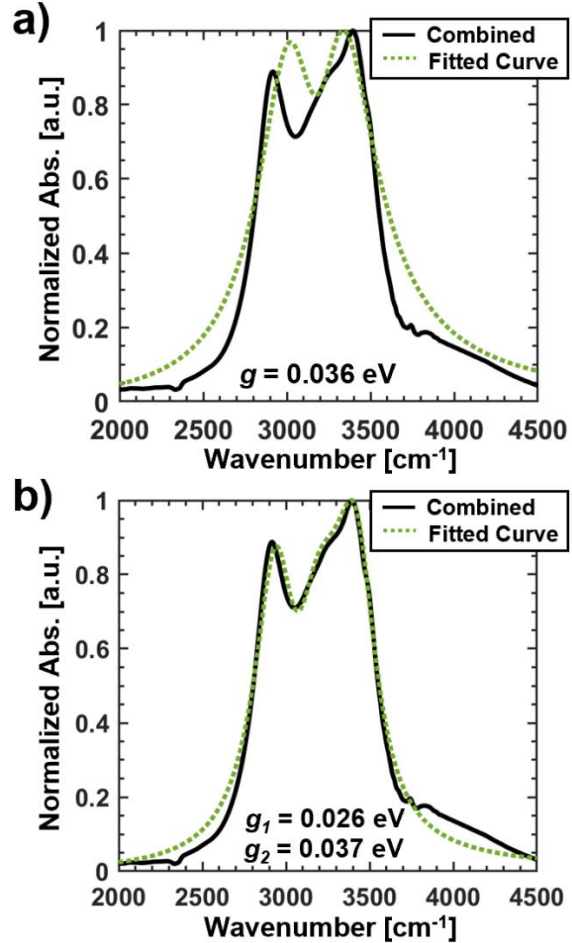

**FIG. 4.** The fits to panel (iv) in Fig. 3 using the (a) two-coupled-oscillator model and (b) three-coupled-oscillator model. The experimental data is shown in black, and the fit from each model is shown by the green dotted line. The coupling strength, $g$, is also indicated.

To account for this more complex coupling interaction, the coupled-oscillator model was extended to allow for simultaneous coupling of two independent vibrational modes to the same plasmonic resonance (see Supplementary Information for details). The utility of this three-coupled oscillator model is for describing situations where two separate orthogonal modes become coupled primarily *via* a strong coupling interaction with a third radiative mode. Similar strategies have been described in other works.[36,38,39] The resulting fits provide a much better description of the data, as illustrated in Fig. 4b. Here, the molecular damping rates and mode frequencies as well as the plasmon damping rates were constrained to the experimentally determined values from the control experiments, and the plasmon frequencies, coupling strengths, and mode amplitudes were left as free fit parameters. Two $g$-values, $g_1$ and $g_2$, correspond to the coupling strength of the symmetric and asymmetric modes, respectively, with the plasmonic mode. Each of the modes will be in the strong-coupling regime if its corresponding coupling strength satisfies:[55,59]

$$g > \frac{1}{4}(\gamma_{pl} + \gamma_{vm})$$

To further ensure that the three-coupled oscillator model accurately describes a system with multiple coupled vibrational modes, we also ran a control experiment with the plasmonic substrate coupled to the isolated, single mode of the C=O vibrational stretch of PMMA. Similar to prior work that studied PMMA coupled to F.P. cavities, we fit the data to a simpler, two-coupled oscillator model.[60] Normalized absorbance spectra of the original plasmon resonance, PMMA on top of an $Al_2O_3$ control substrate, and the spectrum when PMMA was deposited on the plasmonic substrate are shown in Fig. 5a. The shifted plasmon resonance spectral position due to the change in refractive index was obtained by the fit. Again, we observed peak splitting indicative of vibrational strong coupling. The high quality of the fit with a two-coupled oscillator model is shown in Fig. 5b. We found that this system is in the strong coupling regime, since the fitting gave a coupling strength $g = 0.008$ eV $> \frac{1}{4}(\gamma_{pl} + \gamma_{vm})$, where $\gamma_{pl} = 0.013$ eV and $\gamma_{vm} = 0.002$ eV. This study provides further evidence that 1) the two-coupled oscillator model is physically descriptive for a single vibrational mode coupled to a single plasmon resonance; 2) the two-coupled oscillator model cannot be applied to a multi-coupled oscillator system and provide a robust fit; and 3) the three-coupled oscillator model accurately describes the multi-coupled system of the plasmon resonance coupled to two separate vibrational modes, as depicted in Fig 4b.

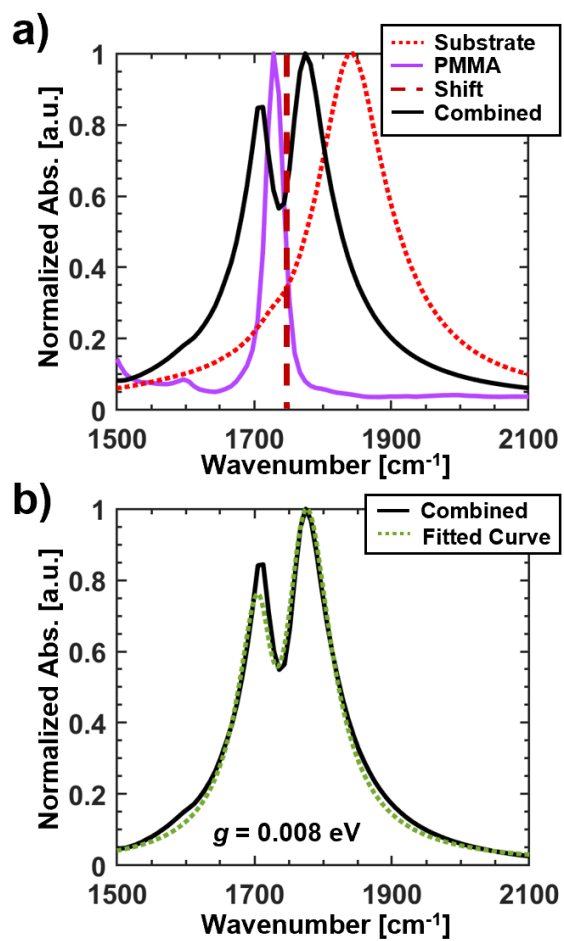

**FIG. 5.** A two-coupled oscillator model applied to the plasmon substrate ($d = 1200$ nm) coupled to a single vibrational mode of PMMA. (a) The spectral absorbance (normalized) of a bare plasmonic substrate (red dashed), PMMA thin film on $Al_2O_3$ with a gold back reflector (purple), and the combined spectra of the PMMA thin film on top of the plasmonic substrate (black). The vertical maroon dashed line shows the shifted plasmon resonance frequency when the surrounding medium is PMMA. The absorbance band of PMMA at 1728 cm$^{-1}$ is the spectral signature of the C=O stretching mode. (b) The two-coupled oscillator model applied to the "combined" curve in (a). The coupling strength is also indicated.

Fig. 6 shows the three-coupled-oscillator model applied to all the spectra in the panels of Fig. 3. Notice that the model in green fits extremely well to the experimental data in black, reproducing well the lineshape even in the case of large plasmon detuning in panels (i) and (vi). Moreover, the high quality of the fits indicates that all the molecules in the thin film on the surface of the substrate that contribute to the far-field spectra participate in coupling when the plasmon resonance is tuned to the vibrational modes, because the three-coupled oscillator model does not account for any uncoupled molecules. That is, the spectra we observe do not appear to be a linear combination of both coupled and uncoupled molecular spectra. This can be explained in part by considering the electric field mode volume of the plasmonic substrate compared with the thin-film thickness. While the $CuSO_4(H_2O)_1$ thin film thickness extends to ~2 µm above the substrate, the electric field enhancement factor $|E/E_0|^2$ produced by the plasmonic substrate extends to similar heights. The volume average of the field enhancement contained at various heights above the substrate and the associated field map was obtained from FEM simulations and are shown in Fig. S3 (Supplementary Information). Although the highest field enhancement resides within 500 nm of the substrate surface, some degree of enhancement also extends up to at least the height of the thin film. This means that all molecular dipoles in the thin film on the substrate surface can be influenced by the enhanced electric field, allowing for some vibrational coupling to occur. Our experiments indicate that any molecules not participating in the coupling interaction, if present, do not contribute significantly to the measured spectra.

Further evidence that the substrate couples to all molecular orientations in the surface film is provided in Fig. S4 (Supplementary Information), which shows that the absolute absorbance of the coupled system is enhanced compared to the uncoupled molecular spectra. This result can be understood as follows: In an uncoupled thin film of molecules, some molecular dipoles will be misaligned with the incident radiation, meaning that the entire ensemble of molecules will not contribute to the measured absorption. However, our angle-independent plasmonic array acts like an antenna, enabling excitation of molecular dipoles along any orientation when the plasmon mode is excited (at normal incidence in our experiments). When the plasmon resonance is tuned to the molecular mode, all molecules within the mode volume of the electric field absorb equally, independent of their dipole orientation. Therefore, the coupled spectra show a greater overall absorbance than the uncoupled system.

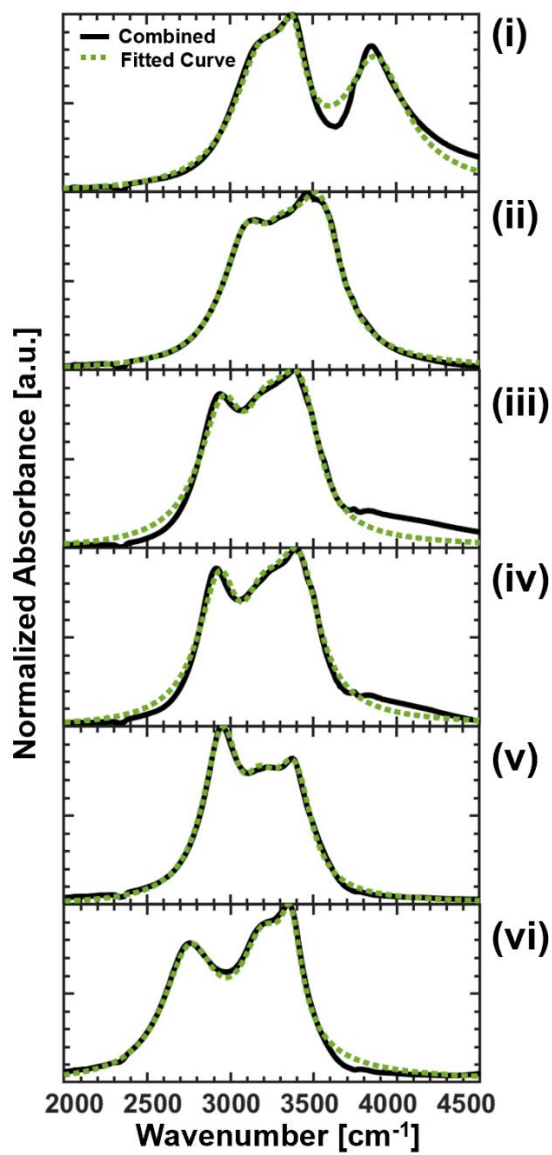

**FIG. 6.** Fits from the three-coupled-oscillator model (green dashed) for all of the "combined" spectra from Fig. 3 of the $CuSO_4(H_2O)_1$ thin film on top of the plasmonic substrate. The data (black) in panels (i)-(vi) correspond directly to data (black) in panels (i)-(vi) of Fig.3, respectively.

The intense fields produced by the plasmonic substrate also give rise to multi-mode coupling. Because the near-field energy density is so large, the plasmonic mode can have large damping rates (large bandwidth) and still couple to both molecular water stretches simultaneously.

A summary of the fitted data to both modes is provided in Table 1. The fit not only determines the coupling strengths but also determines the red-shifted plasmon frequencies obscured by the polariton lineshape. These fitted values are how we report the shifted plasmon resonance positions in Fig. 3 and Fig. 5a. Fig. 7 shows the calculated dispersion relation according to the three-coupled-oscillator model with respect to the plasmon frequency $\omega_{pl}$ using the fitted parameters of Table 1, the average coupling strengths $g_1$ and $g_2$, and the average mode intensities for the cases where the plasmon frequency was in resonance with the water modes [panels (ii) – (v)]. The cases with large plasmon detuning have been omitted because the coupling strength does not contribute significantly when fitting the model. The vertical black dashed lines show the plasmon frequency that corresponds to the experimental data and fits in Fig. 6 panels (ii) – (v). The experimental symmetric and asymmetric water frequencies, $\omega_{vm1}$ and $\omega_{vm2}$, respectively, are shown by the horizontal green dashed lines. As the plasmon tunes through the water modes, hybridized polariton modes and anti-crossing behavior are observed, as expected for the three-coupled oscillator model.

| Panel | $\gamma_{vm1}$ (eV) | $\gamma_{vm2}$ (eV) | $\gamma_{pl}$ (eV) | $\omega_{pl}$ (cm$^{-1}$) | $g_1$ (eV) | $g_2$ (eV) | $g_1 > \frac{1}{4}(\gamma_{pl} + \gamma_{vm1})$ | $g_2 > \frac{1}{4}(\gamma_{pl} + \gamma_{vm2})$ |
|---|---|---|---|---|---|---|---|---|
| i | 0.031 | 0.029 | 0.067 | 3880 | N/A | N/A | N/A | N/A |
| ii | 0.031 | 0.029 | 0.056 | 3274 | 0.026 | 0.038 | Yes | Yes |
| iii | 0.031 | 0.029 | 0.064 | 3161 | 0.025 | 0.037 | Yes | Yes |
| iv | 0.031 | 0.029 | 0.055 | 3129 | 0.027 | 0.034 | Yes | Yes |
| v | 0.031 | 0.029 | 0.046 | 3119 | 0.026 | 0.037 | Yes | Yes |
| vi | 0.031 | 0.029 | 0.049 | 2939 | N/A | N/A | N/A | N/A |

**Table 1.** The experimentally determined values of plasmon $\gamma_{pl}$, symmetric $\gamma_{vm1}$, and asymmetric $\gamma_{vm2}$, mode damping (in blue) based on analysis of uncoupled spectra, as well as the fitted parameters obtained using the three-coupled-oscillator model for analysis of panels (i) – (vi) in Fig. 6. $\omega_{pl}$ is the shifted plasmon position according to the fits, and $g_1$ and $g_2$ are the coupling strengths of the plasmon to the symmetric and asymmetric water stretching modes, respectively. The two columns on the right indicate which of the two vibrational modes are in the strong coupling regime on a panel-by-panel basis (in green). Data labeled N/A (not applicable) cannot be used to determine g-values due to large plasmon detuning.

Upon inspection of Table 1, three observations are clear. (1) All panels where the plasmon is in close resonance with the water modes are in the strong coupling regime, which means that each substrate strongly couples to both vibrational modes. This is also evident from Fig. 7 because the dispersion shows anti-crossing behavior as the plasmon frequency tunes through the vibrational modes. (2) Despite two vibrational modes being present, the coupling drives the plasmon into resonance with the modes. This suggests that when the plasmon shifts into resonance with the water modes, the modes "pull" the plasmon resonance into a similar spectral position, as expected due to the lineshape of the dielectric function near the molecular modes.[45,50,51] (3) The plasmon seems to couple more strongly to the asymmetric water stretch, because $g_2 > g_1$ for all frequencies. This trend suggests intrinsic features of these molecular modes cause the plasmon to couple more strongly with the asymmetric water mode. We hypothesize this is because the asymmetric mode has a smaller damping rate and also a stronger transition dipole moment, as also indicated by the greater peak intensity of that mode in the uncoupled molecular spectrum. We also see that the values of $g_1$ and $g_2$ are relatively constant in all spectra where it is possible to robustly fit for coupling strength, as expected for the coupled oscillator model.[61,62] The variations observed in the coupling strength are likely due to differences in the plasmonic arrays that were fabricated to collect each spectrum, such as variable surface roughness, or changes in the local electric field density when there are different nanodisk diameters.

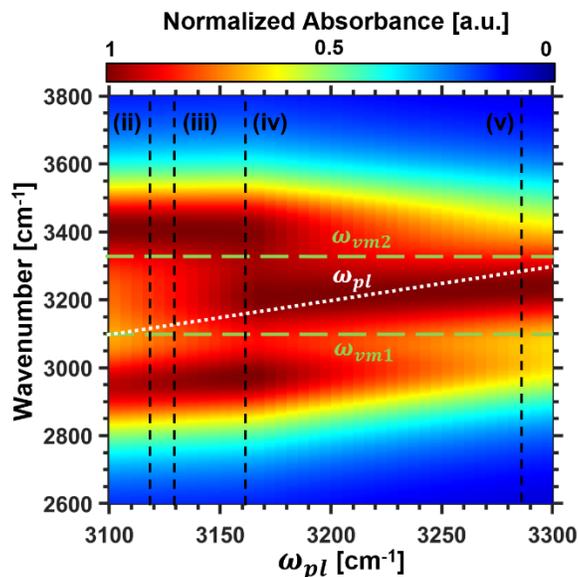

**FIG. 7.** The calculated dispersion based on fits to a three-coupled-oscillator model when tuning the substrate plasmon resonance frequency through the water symmetric $\omega_{vm1}$ and asymmetric $\omega_{vm1}$ stretching modes. The plasmon frequency $\omega_{pl}$ is depicted by the white diagonal line. The vertical black lines show the plasmon frequencies which correspond to panels (ii) – (v) in Fig. 3 and 6.

In conclusion, we have designed and fabricated angle-independent plasmonic nanodisk substrates that simultaneously strongly couple to both the symmetric and asymmetric water stretching modes in a thin film layer of $CuSO_4(H_2O)_1$ at room temperature. By simulating the total hemispherical absorptivity and field distribution of the structure, we found that the plasmon resonance is angle-independent and extends well above the molecular film, suggesting that all molecules in the deposited film strongly couple to the substrate regardless of molecular orientation. Furthermore, we developed a three-coupled-oscillator model that accounted for vibrational strong coupling to two water stretching modes simultaneously in order to analyze the coupling data. Our model confirmed that all molecules on the substrate surface contributing to the far-field signal were strongly coupled to the plasmonic mode. From the model, we also obtained (1) the shifted plasmon resonance positions that were obscured by the strong coupling and (2) the coupling strengths of the plasmonic substrate with the symmetric and asymmetric water stretching modes, $g_1$ and $g_2$, respectively. The magnitude of the coupling strength to either mode as a function of the spectral position of the plasmon resonance was relatively constant, and in general, the plasmon appears to couple more strongly to the asymmetric water stretch. The simultaneous strong coupling of multiple vibrational modes to the same plasmonic resonance results in the coherent exchange of energy between the previously orthogonal molecular vibrational modes, via the substrate interaction, potentially providing a new route for control of chemical behavior.

We believe that the substrate design we demonstrated may benefit future studies of vibrational strong coupling, because the substrate is highly tailorable for targeting specific vibrational modes or coupling bandwidths, and the design eliminates the need for molecules to have the appropriate orientation with respect to the optical mode. In particular, by showing multimode strong coupling to two water vibrational modes, this work could impact future studies of strong coupling in biological systems, since interaction with water is a major feature of biological processes. Further, given that all molecules in the near field of the surface can, in principle, couple to the substrate, it may be possible to achieve even more pronounced modification of the chemical behavior of molecules via vibrational strong coupling, especially if other challenges related to the field inhomogeneity at the substrate surface can be addressed.

*Acknowledgments*

This work was funded by the Air Force Office of Scientific Research under Award No. FA9550-16-1-0154 and TAMU X-Grants. M.S. also acknowledges support from the Welch

Foundation (Grant No. A-1886) and the Gordon and Betty Moore Foundation through Grant No. GBMF6882. M.P. acknowledges support from the National Science Foundation (DMR-1905135). We would like to thank Hayley Brawley for help with the development and troubleshooting of the MATLAB code. We would also like to thank Dr. Nicki Hogan, Ethan Morse, and the TAMU Aggiefab staff for their helpful fabrication advice.

*Supplementary Information*

See supplementary information for the on/off resonance Raman signal, studied plasmonic red shifts, three-coupled-oscillator model, simulated mode volumes, and absolute absorbance spectra of resonant substrates.

*Data Availability Statement*

The data that supports the findings of this study are available within the article [and its supplementary material].

*References*

[1] A. Shalabney, J. George, H. Hiura, J.A. Hutchison, C. Genet, P. Hellwig, and T.W. Ebbesen, Angew. Chem. Int. Ed. **54**, 7971 (2015).
[2] A. Thomas, L. Lethuillier-Karl, K. Nagarajan, R.M.A. Vergauwe, J. George, T. Chervy, A. Shalabney, E. Devaux, C. Genet, J. Moran, and T.W. Ebbesen, Science **363**, 615 (2019).
[3] M. Hertzog, M. Wang, J. Mony, and K. Börjesson, Chem. Soc. Rev. **48**, 937 (2019).
[4] A. Thomas, A. Jayachandran, L. Lethuillier-Karl, R.M.A. Vergauwe, K. Nagarajan, E. Devaux, C. Genet, J. Moran, and T.W. Ebbesen, Nanophotonics **9**, 249 (2020).
[5] C. Climent and J. Feist, Phys. Chem. Chem. Phys. **22**, 23545 (2020).
[6] D. S. Dovzhenko, S. V. Ryabchuk, Y. P. Rakovich, and I. R. Nabiev, Nanoscale **10**, 3589 (2018).
[7] A.D. Dunkelberger, B.T. Spann, K.P. Fears, B.S. Simpkins, and J.C. Owrutsky, Nat. Commun. **7**, 13504 (2016).

[8] B. Xiang, R.F. Ribeiro, M. Du, L. Chen, Z. Yang, J. Wang, J. Yuen-Zhou, and W. Xiong, Science **368**, 665 (2020).

[9] I. Vurgaftman, B.S. Simpkins, A.D. Dunkelberger, and J.C. Owrutsky, J. Phys. Chem. Lett. **11**, 3557 (2020).

[10] M. Hertzog, P. Rudquist, J.A. Hutchison, J. George, T.W. Ebbesen, and K. Börjesson, Chem. Weinh. Bergstr. Ger. **23**, 18166 (2017).

[11] R.F. Ribeiro, L.A. Martínez-Martínez, M. Du, J. Campos-Gonzalez-Angulo, and J. Yuen-Zhou, Chem. Sci. **9**, 6325 (2018).

[12] M. Hertzog and K. Börjesson, ChemPhotoChem **4**, 612 (2020).

[13] H. Memmi, O. Benson, S. Sadofev, and S. Kalusniak, Phys. Rev. Lett. **118**, 126802 (2017).

[14] A. Agrawal, A. Singh, S. Yazdi, A. Singh, G.K. Ong, K. Bustillo, R.W. Johns, E. Ringe, and D.J. Milliron, Nano Lett. **17**, 2611 (2017).

[15] D.K. Gramotnev and S.I. Bozhevolnyi, Nat. Photonics **4**, 83 (2010).

[16] K. Santhosh, O. Bitton, L. Chuntonov, and G. Haran, Nat. Commun. **7**, ncomms11823 (2016).

[17] J.M. Katzen, C. Tserkezis, Q. Cai, L.H. Li, J.M. Kim, G. Lee, G.-R. Yi, W.R. Hendren, E.J.G. Santos, R.M. Bowman, and F. Huang, ACS Appl. Mater. Interfaces **12**, 19866 (2020).

[18] H. Leng, B. Szychowski, M.-C. Daniel, and M. Pelton, Nat. Commun. **9**, 4012 (2018).

[19] M. Kumar, J. Dey, M. Singh Verma, and M. Chandra, Nanoscale **12**, 11612 (2020).

[20] M. Wersäll, J. Cuadra, T.J. Antosiewicz, S. Balci, and T. Shegai, Nano Lett. **17**, 551 (2017).

[21] G. Zengin, M. Wersäll, S. Nilsson, T.J. Antosiewicz, M. Käll, and T. Shegai, Phys. Rev. Lett. **114**, 157401 (2015).

[22] W. Wang, P. Vasa, R. Pomraenke, R. Vogelgesang, A. De Sio, E. Sommer, M. Maiuri, C. Manzoni, G. Cerullo, and C. Lienau, ACS Nano **8**, 1056 (2014).

[23] X. Chen, Y.-H. Chen, J. Qin, D. Zhao, B. Ding, R.J. Blaikie, and M. Qiu, Nano Lett. **17**, 3246 (2017).

[24] A.E. Schlather, N. Large, A.S. Urban, P. Nordlander, and N.J. Halas, Nano Lett. **13**, 3281 (2013).

[25] J.M. Winkler, F.T. Rabouw, A.A. Rossinelli, S.V. Jayanti, K.M. McPeak, D.K. Kim, B. le Feber, F. Prins, and D.J. Norris, Nano Lett. **19**, 108 (2019).

[26] D.G. Baranov, M. Wersäll, J. Cuadra, T.J. Antosiewicz, and T. Shegai, ACS Photonics **5**, 24 (2018).

[27] R. Chikkaraddy, B. de Nijs, F. Benz, S.J. Barrow, O.A. Scherman, E. Rosta, A. Demetriadou, P. Fox, O. Hess, and J.J. Baumberg, Nature **535**, 127 (2016).

[28] L. Efremushkin, M. Sukharev, and A. Salomon, J. Phys. Chem. C **121**, 14819 (2017).

[29] E.A. Muller, B. Pollard, H.A. Bechtel, R. Adato, D. Etezadi, H. Altug, and M.B. Raschke, ACS Photonics **5**, 3594 (2018).

[30] K.S. Menghrajani, G.R. Nash, and W.L. Barnes, ACS Photonics **6**, 2110 (2019).

[31] B.M. Wells, C.M. Roberts, and V.A. Podolskiy, Appl. Phys. Lett. **105**, 161105 (2014).

[32] A.D. Khan and M. Amin, Plasmonics **12**, 257 (2017).

[33] M.N. Abbas, C.-W. Cheng, Y.-C. Chang, M.-H. Shih, H.-H. Chen, and S.-C. Lee, Appl. Phys. Lett. **98**, 121116 (2011).

[34] N. Liu, M. Mesch, T. Weiss, M. Hentschel, and H. Giessen, Nano Lett. **10**, 2342 (2010).

[35] T.D. Dao, K. Chen, S. Ishii, A. Ohi, T. Nabatame, M. Kitajima, and T. Nagao, ACS Photonics **2**, 964 (2015).

[36] W.M. Takele, F. Wackenhut, L. Piatkowski, A.J. Meixner, and J. Waluk, J. Phys. Chem. B **124**, 5709 (2020).

[37] J. George, A. Shalabney, J.A. Hutchison, C. Genet, and T.W. Ebbesen, J. Phys. Chem. Lett. **6**, 1027 (2015).

[38] M. Muallem, A. Palatnik, G.D. Nessim, and Y.R. Tischler, J. Phys. Chem. Lett. **7**, 2002 (2016).

[39] K.S. Menghrajani, H.A. Fernandez, G.R. Nash, and W.L. Barnes, Adv. Opt. Mater. **7**, 1900403 (2019).

[40] I. Gamo, Bull. Chem. Soc. Jpn. **34**, 764 (1961).

[41] R.L. White, Thermochim. Acta **528**, 58 (2012).
[42] R.M.A. Vergauwe, A. Thomas, K. Nagarajan, A. Shalabney, J. George, T. Chervy, M. Seidel, E. Devaux, V. Torbeev, and T.W. Ebbesen, Angew. Chem. Int. Ed. **58**, 15324 (2019).
[43] S. Babar and J.H. Weaver, Appl. Opt. **54**, 477 (2015).
[44] J. Kischkat, S. Peters, B. Gruska, M. Semtsiv, M. Chashnikova, M. Klinkmüller, O. Fedosenko, S. Machulik, A. Aleksandrova, G. Monastyrskyi, Y. Flores, and W.T. Masselink, Appl. Opt. **51**, 6789 (2012).
[45] A.D. Rakić, A.B. Djurišić, J.M. Elazar, and M.L. Majewski, Appl. Opt. **37**, 5271 (1998).
[46] J.R. Howell, R. Siegel, and M.P. Menguc, *Thermal Radiation Heat Transfer*, 5th ed. (CRC Press LLC, Boca Raton, FL, 2010).
[47] N.P. Sergeant, O. Pincon, M. Agrawal, and P. Peumans, Opt. Express **17**, 22800 (2009).
[48] E. Widjaja, H.H. Chong, and M. Tjahjono, J. Raman Spectrosc. **41**, 181 (2010).
[49] Y. Zakharko, A. Graf, and J. Zaumseil, Nano Lett. **16**, 6504 (2016).
[50] A.P. Manuel, A. Kirkey, N. Mahdi, and K. Shankar, J. Mater. Chem. C **7**, 1821 (2019).
[51] P. Törmä and W.L. Barnes, Rep. Prog. Phys. **78**, 013901 (2014).
[52] F. Herrera and J. Owrutsky, J. Chem. Phys. **152**, 100902 (2020).
[53] R.K. Yadav, M.R. Bourgeois, C. Cherqui, X.G. Juarez, W. Wang, T.W. Odom, G.C. Schatz, and J.K. Basu, ACS Nano **14**, 7347 (2020).
[54] O. Bitton, S.N. Gupta, L. Houben, M. Kvapil, V. Křápek, T. Šikola, and G. Haran, Nat. Commun. **11**, 487 (2020).
[55] M. Pelton, S. David Storm, and H. Leng, Nanoscale **11**, 14540 (2019).
[56] S. Rudin and T.L. Reinecke, Phys. Rev. B **59**, 10227 (1999).
[57] X. Wu, S.K. Gray, and M. Pelton, Opt. Express **18**, 23633 (2010).
[58] G. Zengin, G. Johansson, P. Johansson, T.J. Antosiewicz, M. Käll, and T. Shegai, Sci. Rep. **3**, 3074 (2013).
[59] G. Khitrova, H.M. Gibbs, M. Kira, S.W. Koch, and A. Scherer, Nat. Phys. **2**, 81 (2006).
[60] T. Chervy, A. Thomas, E. Akiki, R.M.A. Vergauwe, A. Shalabney, J. George, E. Devaux, J.A. Hutchison, C. Genet, and T.W. Ebbesen, ACS Photonics **5**, 217 (2018).
[61] V. Savona, L.C. Andreani, P. Schwendimann, and A. Quattropani, Solid State Commun. **93**, 733 (1995).
[62] M.S. Skolnick, T.A. Fisher, and D.M. Whittaker, Semicond. Sci. Technol. **13**, 645 (1998).

*Supplementary Information*

*Raman Spectra on and off Substrate*

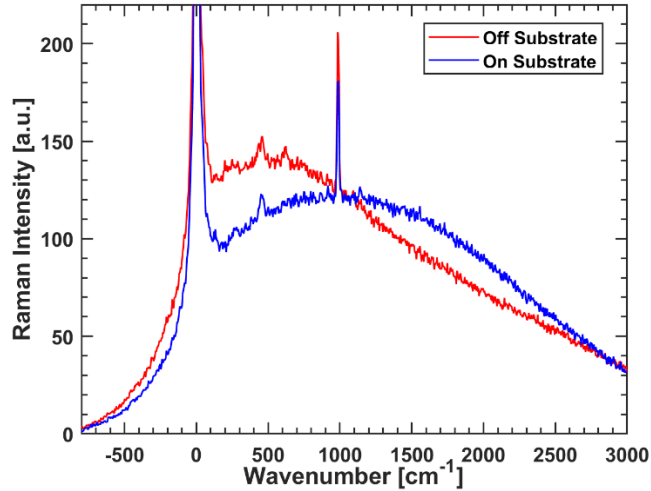

**FIG. S1.** Shows Raman spectra of a thin film of copper sulfate pentahydrate, characterized by the peak at 985 cm$^{-1}$, on versus off a plasmonic substrate that has a resonance around 3300 cm$^{-1}$. No significant plasmonic enhancement of the Raman signal by the substrate is observed.

*Plasmon Redshift with Refractive Index*

As presented in the results and discussion section, the plasmon frequency $\omega_p$ redshifts as a function of the surrounding medium's refractive index. Fig. S2a-c shows the original plasmon frequencies of substrates with different nanodisk diameters when the surrounding medium is air ($n = 1$). When the surrounding medium is CuSO$_4$(H$_2$O)$_1$ the black curves were obtained. In Fig. S2a-c, the black curves clearly still show the plasmon frequency along with the line shape of the CuSO$_4$(H$_2$O)$_1$, indicating that none of the three panels are in the strong coupling regime. Instead, the exact red shift of the plasmon frequency can be obtained, shown by $\Delta\omega_p$. Clearly, the plasmon frequency shifts differently based on the wavenumber, as expected. However, the exact change in red shift is nonlinear as a function of the frequency. Instead, plasmon resonances near the water stretching modes shift more when near the water vibrational modes or less when spectrally isolated.

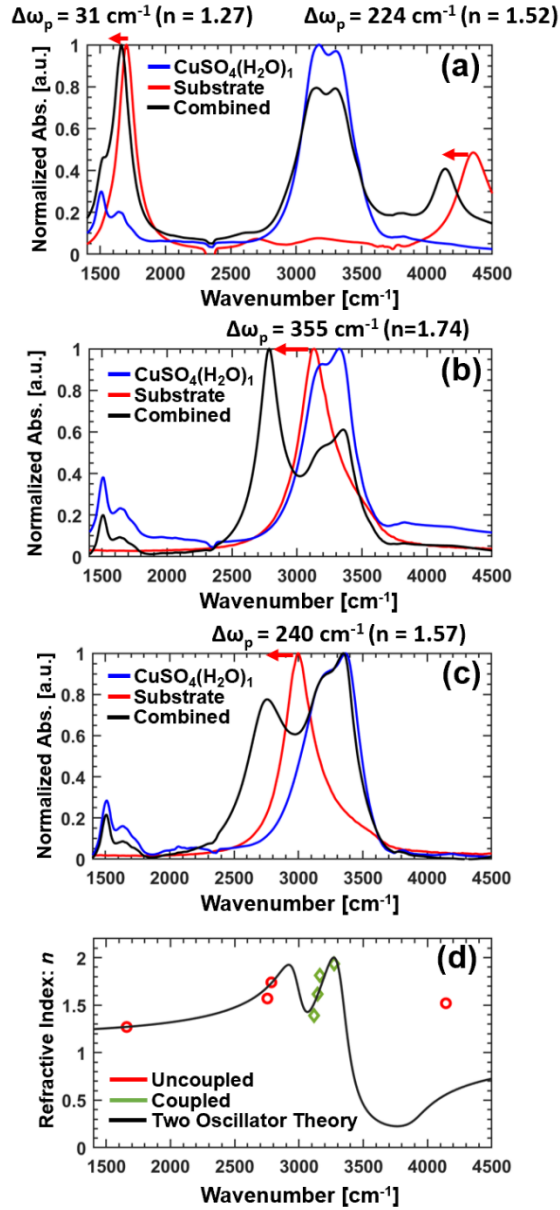

**FIG. S2.** (a-c) Shows how the plasmon frequency redshifts, $\Delta\omega_p$, as a function of the surrounding medium's refractive index at higher and lower wavenumbers. The red curves show the spectra of the bare plasmonic substrate with no $CuSO_4(H_2O)_1$ on top with approximate diameters of (a) $d = 1548$ nm, (b) $d = 736$ nm, and (c) $d = 786$ nm. The blue curves show the spectra of the $CuSO_4(H_2O)_1$ thin

film on top of only $Al_2O_3$. The black curve shows the spectra of the combined $CuSO_4(H_2O)_1$ on top of the nanodisk substrate, which shows how the plasmon frequency shifts as a function of the surrounding medium. (d) Shows the refractive indices we obtained by inspection of panels (a)-(c), which are uncoupled (in red) and from the three-coupled oscillator model in Fig. 6 (ii)-(v) in the case of coupling. The black curve shows the theoretical two oscillator refractive index curve, when we assume that only the water symmetric and asymmetric modes are present.

Furthermore, the exact change in value for the refractive index of $CuSO_4(H_2O)_1$ at different wavenumbers can be obtained using FDTD simulations. Fig. S2a shows a substrate with diameter $d = 1548$ nm, which actually has two plasmon resonances – one primary resonance at 1705 cm$^{-1}$ and a secondary resonance at 4359 cm$^{-1}$. After matching the resonances at $n = 1$, we then slowly swept the refractive index to higher values to match the exact experimental shifts shown in black. The resulting refractive indices we found are shown in Fig. S2a-c for detuned substrates that show no indication of coupling with different nanodisk diameters. The refractive indices at higher wavenumbers are in near-perfect agreement with known values of $CuSO_4(H_2O)_5$.

Fig. S2d shows the refractive indices of copper sulfate in panels (a)-(c), which we discussed above, in red. The green data shows the refractive indices we obtained by matching the FDTD simulations to the redshifted plasmon frequencies in the strongly coupled cases [Fig. 6 (ii)-(v) in the main manuscript], which were obtained using the three-coupled oscillator fit. The black curve shows the theoretical refractive indices determined by the two-oscillator Lorentz model,[1] when we assume that the only modes present in copper sulfate are the symmetric an asymmetric water stretching modes. The data obtained both by inspection (in red) and by the fit (in green) agree very well with the theoretical curve. Any variation in the data, such as the point around 4100 cm$^{-1}$, are likely due to broad, low absorption features of the copper sulfate that are not clearly defined in the literature, as well as potential other modes at higher wavenumbers not captured here spectroscopically.

*Three-coupled-oscillator model*

The coupling between a single transition (electronic or vibrational) and a plasmonic resonance has frequently been modeled by treating the transition and the plasmon as a pair of classical harmonic oscillators, coupled through the near field of the plasmon.[2–4] The corresponding equations of motion are

$$\ddot{\mu}_{pl} + \gamma_{pl}\dot{\mu}_{pl} + \omega_{pl}^2\mu_{pl} = F_{pl} + g(\omega_{pl}d_{pl}/d_{vm})\mu_{vm}$$

$$\ddot{\mu}_{vm} + \gamma_{vm}\dot{\mu}_{vm} + \omega_{vm}^2\mu_{vm} = g(\omega_{vm}d_{vm}/d_{pl})\mu_{pl}$$

where $\mu_{pl}$ is the oscillating dipole of the plasmon; $\omega_{pl}$ is the plasmon resonance frequency; $\gamma_{pl}$ is the plasmon linewidth; $d_{pl}$ is the transition dipole moment of the plasmon; $\mu_{vm}, \omega_{vm}, \gamma_{vm}$, and $d_{vm}$ are the corresponding quantities for the vibrational mode; $g$ is the coupling strength; and $F_o$ is a driving term representing the force of an external optical field on the plasmon. Here, it is assumed that the plasmon cross-section is much larger than that of the vibrational mode, so that direct driving of the vibration by the external field can be ignored. Solving the equations of motion for an external field of form $F_{pl} = F_{pl}e^{i\omega t}$, we obtain

$$\mu_{pl} = \frac{F_{pl}(\omega_{vm}^2 - \omega^2 - i\omega\gamma_{vm})}{(\omega_{vm}^2 - \omega^2 - i\omega\gamma_{vm})(\omega_{pl}^2 - \omega^2 - i\omega\gamma_{pl}) - \omega_{vm}\omega_{pl}g^2}$$

The extinction cross-section can then be found according to $\sigma_{ext} \propto F_{pl}\omega\Im[\mu_{pl}]$.

To describe the situation in the present study, we extend the model to include two vibrational modes, described by quantities $\mu_{vm1}, \omega_{vm1}, \gamma_{vm1}, d_{vm1}$ and $\mu_{vm2}, \omega_{vm2}, \gamma_{vm2}, d_{vm2}$, each coupled independently to the same plasmon mode, with corresponding coupling strengths $g_1$ and $g_2$. We also allow for direct absorption of the incident field by the vibrational modes by including driving terms $F_{vm1} = F_{vm1}e^{i\omega t}$ and $F_{vm2} = F_{vm2}e^{i\omega t}$. The vibrational modes are assumed not to be coupled to each other. The resulting equations of motion are

$$\ddot{\mu}_{pl} + \gamma_{pl}\dot{\mu}_{pl} + \omega_{pl}^2\mu_{pl} = F_{pl} + g_1(\omega_{pl}d_{pl}/d_{vm1})\mu_{vm1} + g_2(\omega_{pl}d_{pl}/d_{vm2})\mu_{vm2}$$

$$\ddot{\mu}_{vm1} + \gamma_{vm1}\dot{\mu}_{vm1} + \omega_{vm1}^2\mu_{vm1} = F_{vm1} + g(\omega_{vm1}d_{vm1}/d_{pl})\mu_{pl}$$

$$\ddot{\mu}_{vm2} + \gamma_{vm2}\dot{\mu}_{vm2} + \omega_{vm2}^2\mu_{vm2} = F_{vm2} + g(\omega_{vm2}d_{vm2}/d_{pl})\mu_{pl}$$

Solving these coupled equations gives

$$x_{pl} = \frac{F_{pl}}{A}\Big[(\omega_{vm1}^2 - \omega^2 - i\omega\gamma_{vm1})(\omega_{vm2}^2 - \omega^2 - i\omega\gamma_{vm2}) + g_1\omega_{pl}(\omega_{vm2}^2 - \omega^2 - i\omega\gamma_{vm2})$$
$$+ g_2\omega_{pl}(\omega_{vm1}^2 - \omega^2 - i\omega\gamma_{vm1})\Big]$$

$$x_{vm1} = \frac{F_{vm1}}{A}\Big[(\omega_{pl}^2 - \omega^2 - i\omega\gamma_{pl})(\omega_{vm2}^2 - \omega^2 - i\omega\gamma_{vm2}) + g_1\omega_{vm1}(\omega_{vm2}^2 - \omega^2 - i\omega\gamma_{vm2})$$
$$+ g_1g_2(\omega_{vm1} - \omega_{vm2})\omega_{pl}\Big]$$

$$x_{vm2} = \frac{F_{vm2}}{A}\Big[(\omega_{pl}^2 - \omega^2 - i\omega\gamma_{pl})(\omega_{vm1}^2 - \omega^2 - i\omega\gamma_{vm1}) + g_2\omega_{vm2}(\omega_{vm1}^2 - \omega^2 - i\omega\gamma_{vm1})$$
$$+ g_1g_2(\omega_{vm2} - \omega_{vm1})\omega_{pl}\Big]$$

where

$$A \equiv (\omega_{vm1}^2 - \omega^2 - i\omega\gamma_{vm1})(\omega_{vm2}^2 - \omega^2 - i\omega\gamma_{vm2})(\omega_{pl}^2 - \omega^2 - i\omega\gamma_{pl}) - \omega_{vm1}\omega_{pl}g_1^2(\omega_{vm2}^2 - \omega^2 - i\omega\gamma_{vm2})$$
$$- \omega_{vm2}\omega_{pl}g_2^2(\omega_{vm1}^2 - \omega^2 - i\omega\gamma_{vm1})$$

The extinction cross-section is now given by

$$\sigma_{ext} \propto F_{pl}\omega\Im[x_{pl}] + F_{vm1}\omega\Im[x_{vm1}] + F_{vm2}\omega\Im[x_{vm2}]$$

*Plasmon Enhanced Electric Field Mode Volume*

The electric field enhancement factor $|E/E_0|^2$ surrounding the plasmonic substrate is a measurement of the localized electric field compared to the incident electric field.[5] It was necessary to calculate the enhancement factor in our simulations in order to determine where the highest field concentration would be above the nanodisk substrate. Only molecules within the enhanced electric field mode volume will couple to the plasmonic substrate because this is the region where the highest field intensity will be, allowing the molecules to resonantly exchange energy with the substrate, creating the conditions for a strongly coupled system.[6]

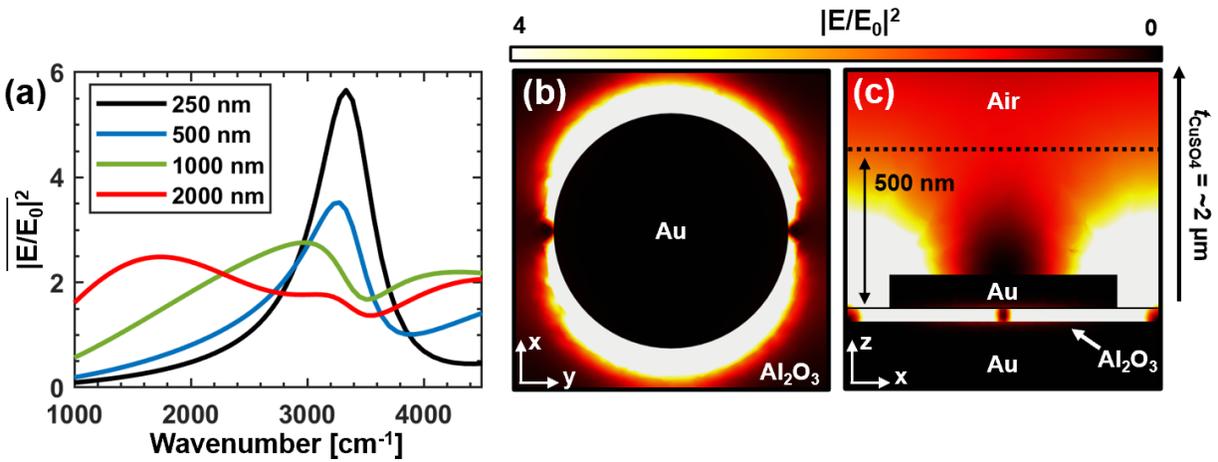

**FIG. S3.** FEM simulation of the electric field enhancement factor $|E/E_0|^2$ of the $d$ = 680 nm nanodisk substrate. (a) Shows the volume averaged enhancement factor at various heights above the surface of the substrate. (b) An electric field map of the nanodisk substrate as viewed at normal to the substrate surface, down the z-axis. (c) An electric field map of the nanodisk substrate as viewed from the side, looking down the y-axis.

To determine the mode volume, a simulation was conducted of the frequency versus $|E/E_0|^2$ in COMSOL with a nanodisk diameter $d$ = 680 nm. Fig. S3a shows the volume averaged

$|E/E_0|^2$ plot at different heights above the substrate, which takes the volume-averaged enhancement factor contained between the specified height and the substrate surface. From the figure, it is clear that the highest electric field concentration at the desired resonance of 3333 cm$^{-1}$ is near the substrate surface, indicated by the largest $|E/E_0|^2$ value being obtained within 250 nm of the surface. As the integration region extends higher above the substrate surface, $|E/E_0|^2$ decreases; however, 2 µm above the substrate still shows field enhancement.

Fig. S3b shows the electric field distribution around a single nanodisk as viewed from the top of the substrate under linearly polarized light in the x-direction. We applied periodic boundary conditions to simulate an infinite array of nanodisks. Let it be noted that the volume averaged field enhancement factor would likely increase under unpolarized light due to the contribution of both S and P polarizations; however, the plasmon resonant frequency would not change as a function of the polarization, as indicated by the angle independence of Fig. 2c in the main body of the text. Fig. S3c shows the electric field distribution as viewed from the side.

*Substrate Enhanced Total Absorbance*

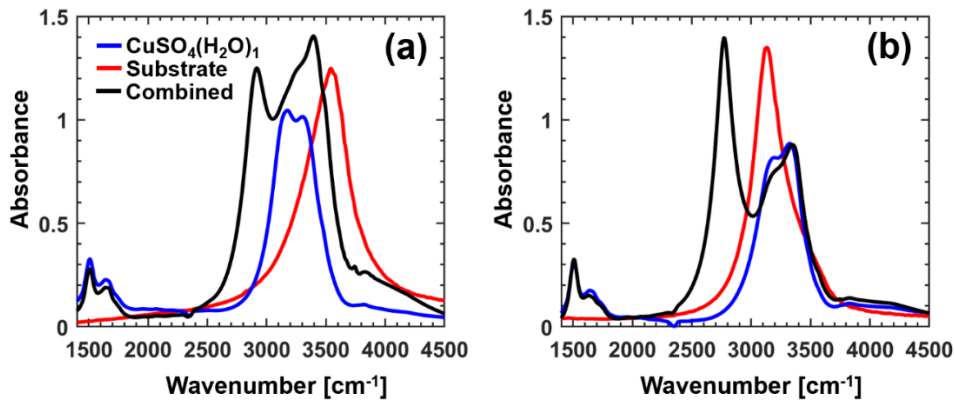

**FIG. S4.** Absorbance spectra collected with an FTIR microscope showing a coupled substrate (a) and uncoupled substrate (b).

Fig. S4 shows that the absorbance [$Log_{10}$(absorptance)] of a thin film of molecules coupled to the angle-independent plasmonic substrate is higher than an uncoupled system. This occurs because the plasmonic substrate acts as an antenna to allow for more light absorption by the molecules. When the substrate is not present, part of the molecular dipoles are misaligned with the incident electric field. However, when the substrate is present, the angle-independent plasmonic mode volume enables interaction with all molecules, no matter the orientation, meaning that more molecules can absorb the light. This is in comparison to Fig. S4b, which shows a detuned plasmonic substrate from the molecular mode resonances, showing no increased absorbance by the molecular modes.

*References*

[1] A.D. Rakić, A.B. Djurišić, J.M. Elazar, and M.L. Majewski, Appl. Opt., AO **37**, 5271 (1998).
[2] M. Pelton, S. David Storm, and H. Leng, Nanoscale **11**, 14540 (2019).
[3] X. Wu, S.K. Gray, and M. Pelton, Opt. Express, OE **18**, 23633 (2010).
[4] S. Rudin and T.L. Reinecke, Phys. Rev. B **59**, 10227 (1999).
[5] J.A. Schuller, E.S. Barnard, W. Cai, Y.C. Jun, J.S. White, and M.L. Brongersma, Nature Materials **9**, 193 (2010).
[6] D. S. Dovzhenko, S. V. Ryabchuk, Y. P. Rakovich, and I. R. Nabiev, Nanoscale **10**, 3589 (2018).